\pdfoutput=1  
\documentclass{svmult}

\usepackage{verbatim}
\usepackage[caption=false, font=footnotesize]{subfig}

\usepackage{mathptmx}       
\usepackage{helvet}         
\usepackage{courier}        
\usepackage{type1cm}        
\usepackage{makeidx}         
\usepackage{graphicx}        
\usepackage{multicol}        
\usepackage[bottom]{footmisc}

\begin{document}

\title*{Transformation Networks: How Innovation and the Availability of Technology can Increase Economic Performance}

\author{Christopher D. Hollander and Ivan Garibay}
\institute{Christopher D. Hollander \at University of Central Florida, \email{chris.hollander@gmail.com}
\and Ivan Garibay \at University of Central Florida, \email{Ivan.Garibay@ucf.edu}}

\maketitle

\abstract{A transformation network describes how one set of resources can be transformed into another via technological processes. Transformation networks in economics are useful because they can highlight areas for future innovations, both in terms of new products, new production techniques, or better efficiency. They also make it easy to detect areas where an economy might be fragile. In this paper, we use computational simulations to investigate how the density of a transformation network affects the economic performance, as measured by the gross domestic product (GDP), of an artificial economy. Our results show that on average, the GDP of our economy increases as the density of the transformation network increases. We also find that while the average performance increases, the maximum possible performance decreases and the minimum possible performance increases. }

\section{Introduction}

\begin{figure}
\sidecaption
\centering
\includegraphics[width=70mm]{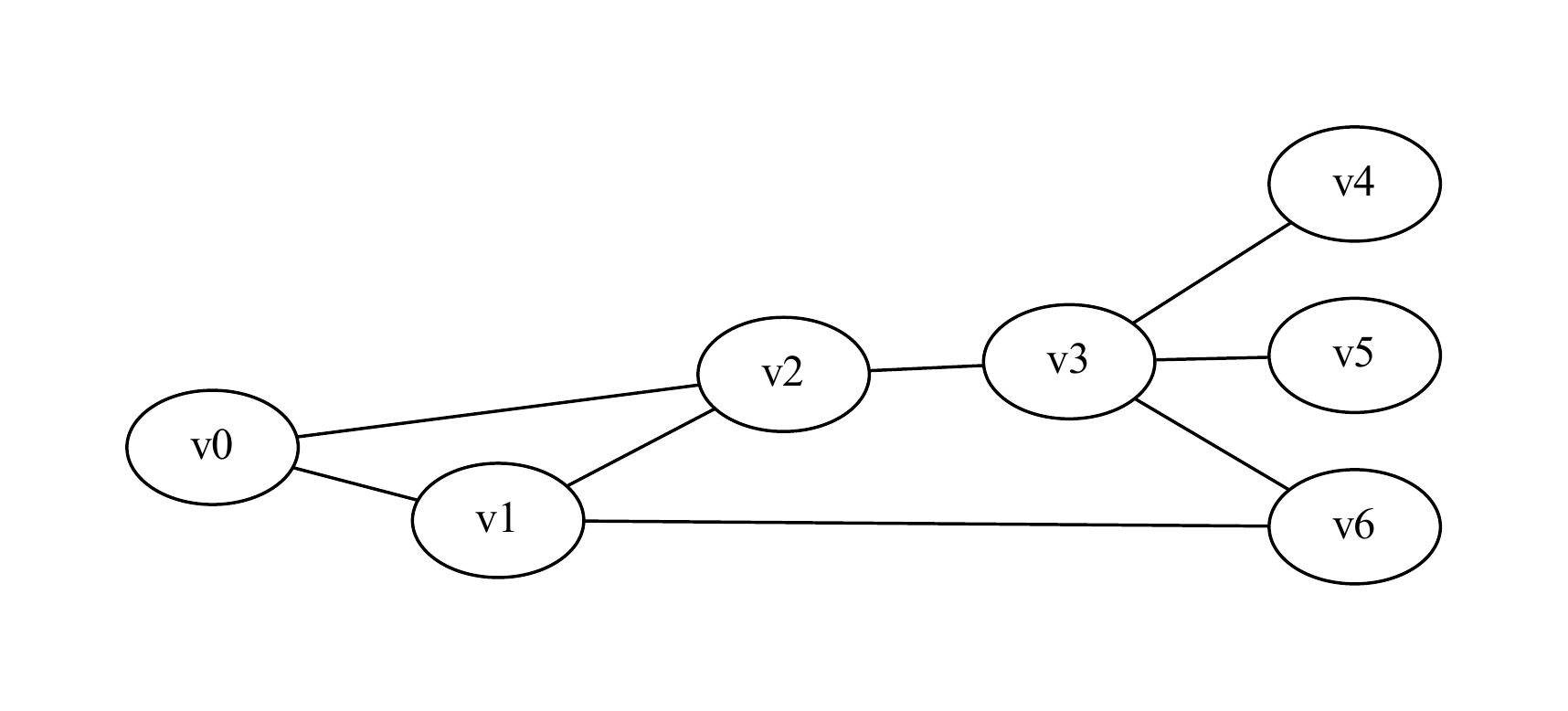}
\caption{A sample undirected network of 7 nodes and 8 edges.}
\label{fig:simple_network}
\end{figure}

Networks have become an increasingly important topic in the study of economic systems \cite{schweitzerf2009, konigm2009}. A network is a collection of objects, called \emph{nodes}, that are joined together by one or more connections, called \emph{edges} \cite{newmanm2010} (Figure \ref{fig:simple_network}). In \emph{economic networks}, the nodes represent economic entities, such as firms or households, and the edges represent economic relationships, such as trade, ownership, or credit-debit relationships.

Economic networks are typically studied through either the analysis of existing socioeconomic data or \emph{in silico} through simulation. However, because it can be difficult to get the required data, simulation is being relied upon more and more to make theoretical advancements.

One particular type of simulation that is often used to study network behavior is \emph{computational agent-based simulation}. Computational agent-based simulations of economic networks are built on top of computational agent-based models. In economics, computational agent-based models differ from the more general ``agent-based models'' in that the computational agents used are non-representative and exist as small, often autonomous and intelligent, computer programs that interact with each other in a virtual environment \cite{colanderd2008, lebaronb2008, tesfatsionl2006}. 

The agents used in computational agent-based simulations are typically heterogeneous, each with its own preferences and rules of behavior. This heterogeneity provides a natural way for agents to determine how they must respond in an arbitrary environment, thus making it straightforward to determine the aggregate behavior of an economic system in non-homogeneous conditions. The agents are also decentralized, in that sense that no one agent controls all the others, and exist in an environment that represents either an explicit space or a set of relationships. Interactions between agents are limited to their local area, either by distance or the existence of a relationship. Often times, the agents possess a degree of intelligence and autonomy \cite{tesfatsionl2008}, but current efforts are far behind the state of the art in computer science (e.g. \cite{russells2009, wooldridgem2009}).

Computational agent-based simulations can involve tens to hundreds of agents. Systems with multiple interacting agents are referred to as \emph{multi-agent systems} (\emph{MAS}).  Multi-agent systems are typically studied at both the micro and macro level. At the micro level, the concern is on the rules that govern the behavior of each individual agent. At the macro level, concern is on the patterns and regularities of behavior that ``emerge'' in the agent population as a result of the agents obeying the micro level rules \cite{arthurb2006, axtellr2006, chensh2011a, cristellic2011, epsteinj2006, gattid2010}\footnote{For a more in depth explanation of agent-based modeling and multi-agent systems, see \cite{epsteinj2006, wooldridgem2009}.}.

When used to simulate economic networks, computational agent-based simulations focus on the relationships and interaction patterns of an agent population that is driven by an agent-based model. Each node in the network is represented by an agent, and each edge represents a type of interaction between two agents. Multiple economic networks can exist on top of a single population of agents. The edges in the networks are created by the actions of the agents, and at the same time also help regulate the behavior of the agents. Many of the networks studied in the existing research have traditionally been static, but there is an increase in the investigation of dynamic networks as the domain becomes better defined and better understood.

The current research on economic networks tends to focus on the more obvious economic relationships, such as ownership, labor markets, diffusion, organizational modeling, R\&D collaborations, trade networks or supply chain networks \cite{konigm2009}. However, these are not the only networks that exist in economic systems. In this paper, we define the concept of a transformation network and use it to study how innovation can increase economic performance and lead to periods of economic growth.

\begin{figure}
\sidecaption
\centering
\includegraphics[width=110mm]{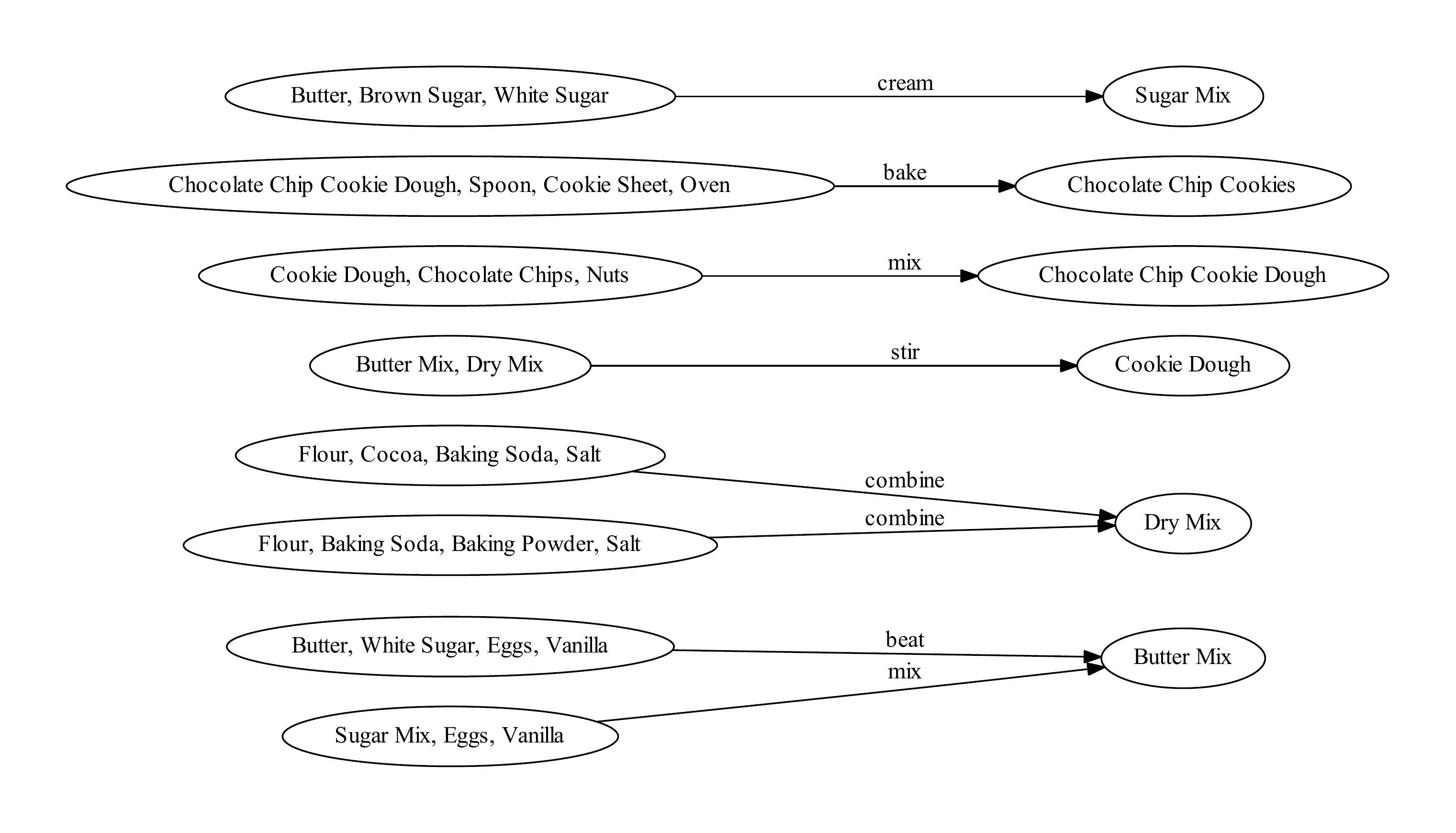}
\caption{A simplified resource transformation network for two chocolate chip cookie recipes that vary by the components used to create the dry and butter mixes.}
\label{fig:res_tran_network}
\end{figure}

A transformation network describes how one set of resources can be transformed into another, such as how a saw can be used to transform trees into lumber and paper, or how amino acids are transformed into proteins under biological processes. In a transformation network, the nodes represent resource sets and a directed edge connects two resource sets if one can be directly transformed into the other. These networks can be static or dynamic. In dynamic transformation networks, the process of adding new nodes or edges to the system over time can be interpreted as invention, innovation, adaptation, or evolution. Figure \ref{fig:res_tran_network} represents a simple static transformation network describing how chocolate chip cookies are produced from a basic set of resources and a set of technologies. Of particular interest in figure \ref{fig:res_tran_network} is that multiple resource sets can be used to produce the same intermediate resources, dry mix and butter mix, under the same or different transformation processes.

When applied to economies, transformation networks describe how natural and man-made resources can be turned into products through technological processes. They can be used to help identify potential areas for innovation, which exist whenever two unconnected nodes can be connected through a new or existing technology. In this way, transformation networks can be helpful to entrepreneurs and corporations seeking to expand their operations to new, but related, areas. In addition, analysis of transformation networks can help identify where the system may be fragile; e.g. if multiple outputs are the product of a single input, the removal of that input could cause catastrophic failure \cite{thadakamallah2004}. The knowledge gained from applying transformation networks to economies also has the potential to help in economic theorizing, particularly with regard to developing new models and explanations of economic growth.

In this paper, we use a computational agent-based simulation to investigate the relationship between transformation networks and the economic performance of a simple artificial economy. In particular, we seek to answer the question, "do transformation networks with more edges yield better economic performance as measured by GDP?" The notion behind this question is that the density of a transformation network is related to the previous amount of innovation that has occurred in the network, and the current amount of technology available in the economy. We begin by introducing the concept of a transformation network. We then outline the experiments covered in this paper and briefly discuss the simulation model used to conduct those experiments. Finally, we describe our experimental results and place them in an applied context.

\section{Transformation Networks}

Transformation networks describe how sets of resources can be transformed from one to another. A transformation network (figure \ref{fig:res_tran_network}) is a simple directed graph $G = (V, E)$, composed of a node set $V$, an edge set $E = \{(v_0, v_1) : v_0, v_1 \in V\}$, and an associated a transformation function $T(u) = v$ where $T : V \rightarrow V$. Each node represents a combination of resources available to the system. An edge connects two nodes $v_0$ and $v_1$ if and only if $v_1$ can be obtained by a single transformation action on $v_0$, i.e. $T(v_0) = v_1$. Edges may have a weight associated with them that represents an aggregate measure of the system, such as the number of entities in the system capable of transforming $v_0$ into $v_1$, or the average cost of performing the associated transformation. The total number of nodes in the system depends on the number of resources. In particular, for $R$ resources the number of total nodes is equal to $2^R$ and each node represents an element of the power set of $R$. 

Transformation networks are similar to production chains, supply chain networks and metabolic networks in that they model the flow of resources through a system. In particular, they describe how resources can be transformed as they are processed by the associated system. When applied to economics, weighted edges can represent the cost of a transformation, and transformation networks can be used to identify cheaper methods of production by calculating the minimum flow over the network. Missing edges can be used to identify potential innovations. The existence of cycles suggests that part of the system may not depend on sustained input for continued operation.

As a first step towards investigating the applications of transformation networks to economics, and to motivate future research in this area, this paper focuses on the relationship between the density of a transformation network and the economic performance of a simple artificial economy, as measured by the Gross Domestic Product (GDP). We hypothesize that as the density of the economy's transformation network increases, the GDP will also increase. The basis for this hypothesis centers around the notion that as the edge density of the transformation network increases, there are more direct ways to produce one resource from another. This yields more overall options for production, thereby increasing the diversity of the system. The level of diversity in an economy is important because it has been shown to have a positive effect on economic performance \cite{pages2010, weitzmanm1998}. In addition, it has been shown that the longer a production or supply chain is the more fragile it is to disruption \cite{thadakamallah2004}. Thus, if a transformation network is dense due to the diversity of transformation rules, it is much more likely to be composed of shorter chains lengths, thereby increasing the robustness of the system. This increase in robustness should be followed by an associated increase in the GDP as compare to sparse transformation networks.

\section{Experimental Setup}

We use a computational agent-based simulation developed with MASON \cite{lukes2005} to investigate the relationship between the density of the transformation network and GDP of a simple artificial economy. 

The economy that we consider consists of a fixed population of immortal agents, $A$, and a fixed set of resources, $R$, in an environment, $E$. Each agent has a single transformation rule, $t \in T$, that represents technological knowledge and a variable amount of wealth, $w_a : a \in A$, that it can use to buy a resource needed as input for its transformation rule. The exact transformation rule an agent possesses is determined randomly at the start of the simulation. Each resource is represented as a b-bit binary string $R \in \{0,1\}_b$ and has a price $p_r : r \in R$. For this particular experiment, the pricing model has been simplified so that the price of each resource is fixed and identical.

Trade between agents occurs at each timestep, $\tau$, when one agent, the \emph{buyer}, buys resources from another agent, the \emph{seller}, according to an economic strategy, $S$ . In our current artificial economy the trade network of the agent population is completely connected and the trade activity of agents is driven by a baseline strategy. Under this strategy an agent will only attempt buy resources if it does not already have the resources required by its transformation rule, for example an agent with the rule $T(01) = 11$ will only buy the resource $01$ if it does not currently have any. Furthermore, a buyer will only attempt to buy from the seller if that agent is selling the required resources for a price $p_r < w_i$. Under this strategy, potential sellers are randomly selected from the entire population at the start of each turn. In addition, each resource is priced at the same fixed value and there are no costs associated with any agent behavior outside of trade. 

Each time one agent successfully trades with another, the value of the GDP increases by the value of the transaction. In this manner, the GDP maintains its macroeconomic definition as a measure of the final value of goods and services produced within the system, $GDP = C + G + I + NX - IM$ with $G = 0, I = 0, NX = 0, IM = 0$, where C is the consumer spending, G is the government spending, I is investment, NX is the net exports, and IM is the net imports \cite{krugmanp2005}. Because agents are immortal and there are no imports and no exports, the artificial economy that we consider is a closed system.

For the particular experiment studied in this paper, we focus on how the GDP of the system responds to changes in the network density, $\rho$, under various uniform distributions on the set of transformation rules, $T$. We also account for scaling phenomenon that might occur given the total number of agents, $|A|$. We expect that GDP will increase monotonically as both $|A|$ and $\rho$ increase.

To obtain statistically significant measurements we run each experimental configuration for 20 replications and 1,000 time steps. This is sufficient because both the underlying trade network and the transformation network are static. An experimental configuration consists of a single choice from the distribution of transformation rules and a constant number of nodes. We consider only uniform distributions of the transformation rules over four resources, $00$, $01$, $10$, and $11$ with the additional constraint that $T(v) \neq v$ (there are no self-loops ). This yields a total of 12  transformation rules to be considered and 4095 unique combinations of a directed graph structure over the four resources. Because the network is directed, its density is computed as $\rho = \frac{|E|}{|V|(|V| - 1)}$. For each distribution choice, we vary the number of agents over the set $\{25, 50\}$ for a total of 8190 data points. Each data point represents the mean GDP value under the associated experimental configuration after disregarding the first 10 time steps in order to allow the simulation to settle. In addition to this primary experiment, another test was done on the undirected combinations of the rule set using agent counts of $\{25, 50, 75, 100\}$. This experiment was conducted in order to verify scaling over population size. The undirected experiment was also repeated with a set of 5 nodes in the transformation network to verify the consistency of the results. 

We only consider small numbers of nodes for this experiment because we wanted to examine all possible edge configurations of the transformation network, and the number of edge configurations grows combinatorially. With 4 nodes, there can be up to 4095 simple directed graphs. With 5 nodes, the number of possible graphs exceeds 1 million\footnote{The number of possible graphs on $n$ nodes is given by summing all combinations of simple graphs with $k$ edges, where $k \in [1 , n ( n - 1)] $. In other words, the total number of possible directed graphs on $n$ nodes is given by $\sum_{k=1}^{n (n-1)} {n(n-1) \choose{k}}$ }. This makes the combinatorial investigation of large graphs with hundreds or thousands of nodes computationally prohibitive.

\section{Results}

Our experiment yields two important results.

First, as the density of the transformation network, $\rho$, increases so too does the average of the average GDP of the underlying artificial economy (figure \ref{fig:plots_a}). Furthermore, the increase in GDP is monotonic and appears to scale linearly with the number of agents in the population. This scaling occurs in both our primary experiment and our secondary experiments on undirected resource transformation networks, but it is most likely a result of our specific underlying economic model.

Second, we find that as the density of the transformation network increases, the maximum GDP value of each configuration group (all graphs with 1 edge, all graphs with 2 edges, etc.) decreases monotonically. At the same time, the minimum GDP value of each configuration group increases. This behavior is illustrated in figure \ref{fig:plots_b}. Of particular interest is how there appears to be a phase transition in the minimum GDP value when $\rho > 0.66$. On our four node network, this occurs when there are 9 out of 12 edges. When this occurs, the transformation network is guaranteed to have at least three cycles on its four nodes. These results imply that, as one would expect, the specific edge structure of the network can play a significant part in the economic performance of a system. Networks without any cycles tend to yield positive economic performance only until all resources have reached their final state, while networks with cycles can reuse resources. This result also suggests that there should be a difference in economic performance if the underlying transformation network is scale-free or small world, but that analysis is outside the scope of this current paper. 

\begin{figure}
\centering

\subfloat[]{\label{fig:plots_a}\includegraphics[width=58mm]{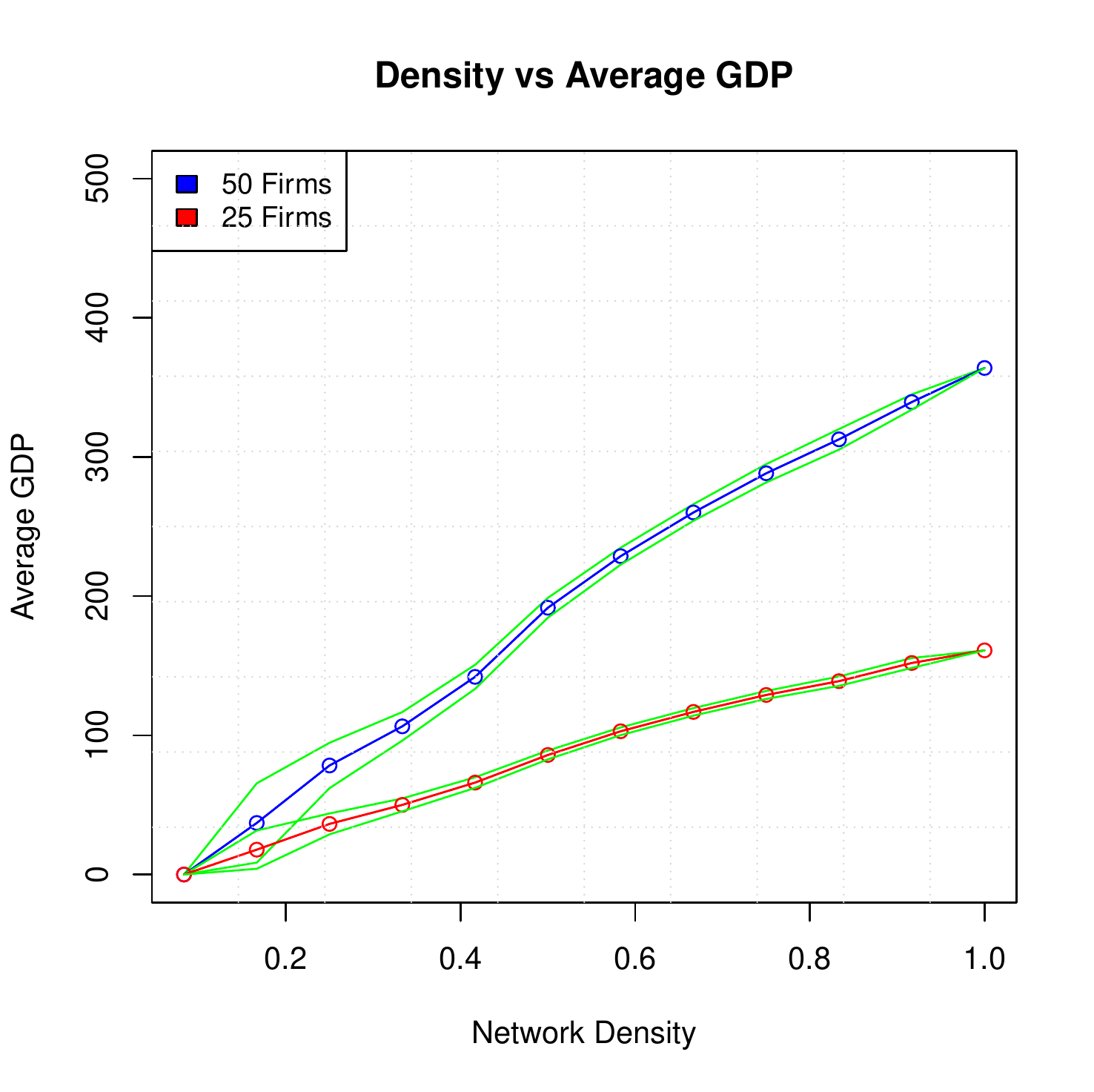}}
\hfill
\subfloat[]{\label{fig:plots_b}\includegraphics[width=58mm]{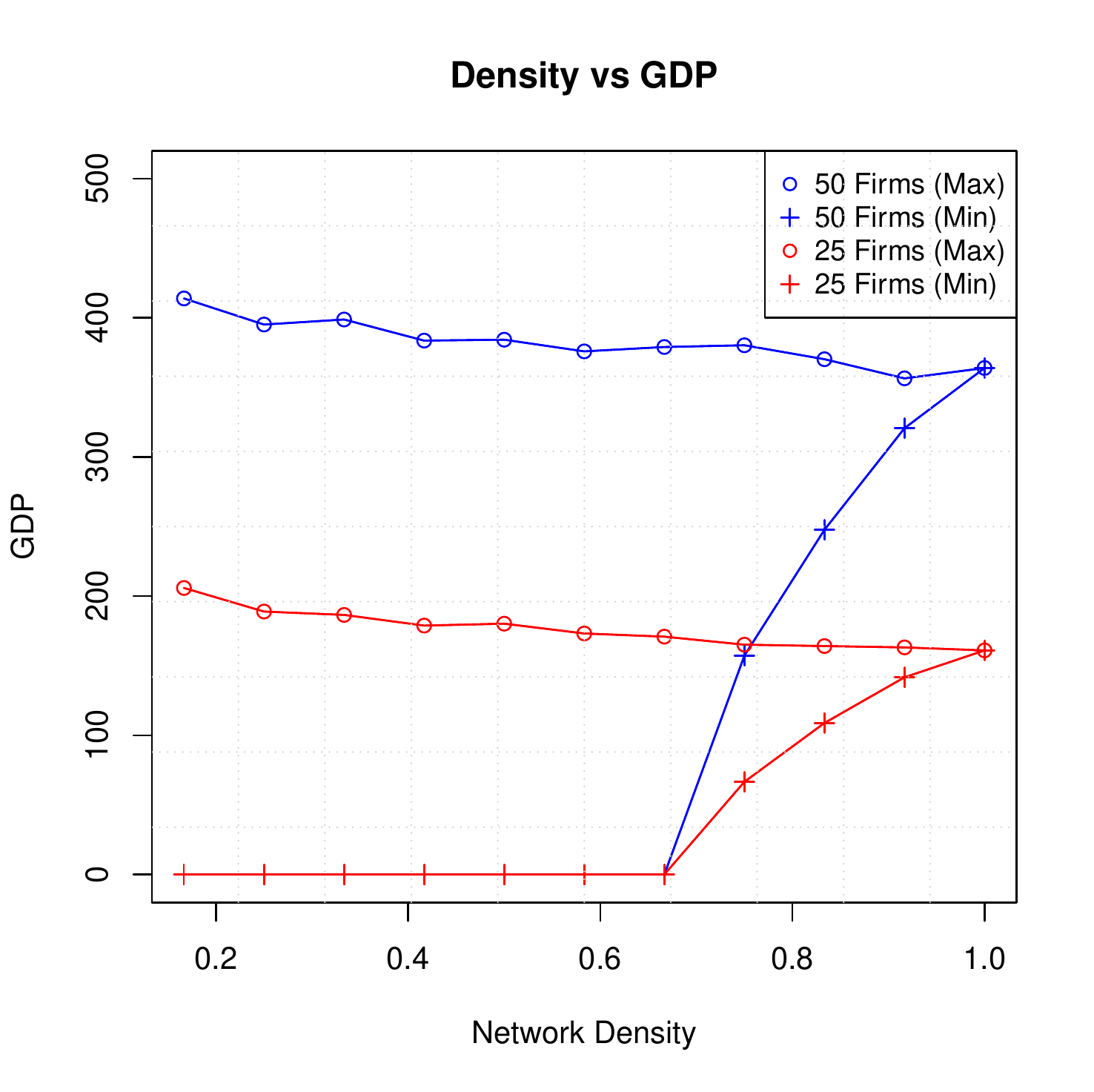}}

\caption{ (a) The average of the average GDP of two populations plotted against the density of the transformation network. Both populations use the same transformation network. The outline around each line represents data corresponding to the 95\% confidence interval. (b) Minimum and maximum GDP values between a population of 25 agents and 50 agents for varying density values.}
\label{fig:plots0}
\end{figure}

\begin{figure}
\centering
\includegraphics[width=110mm]{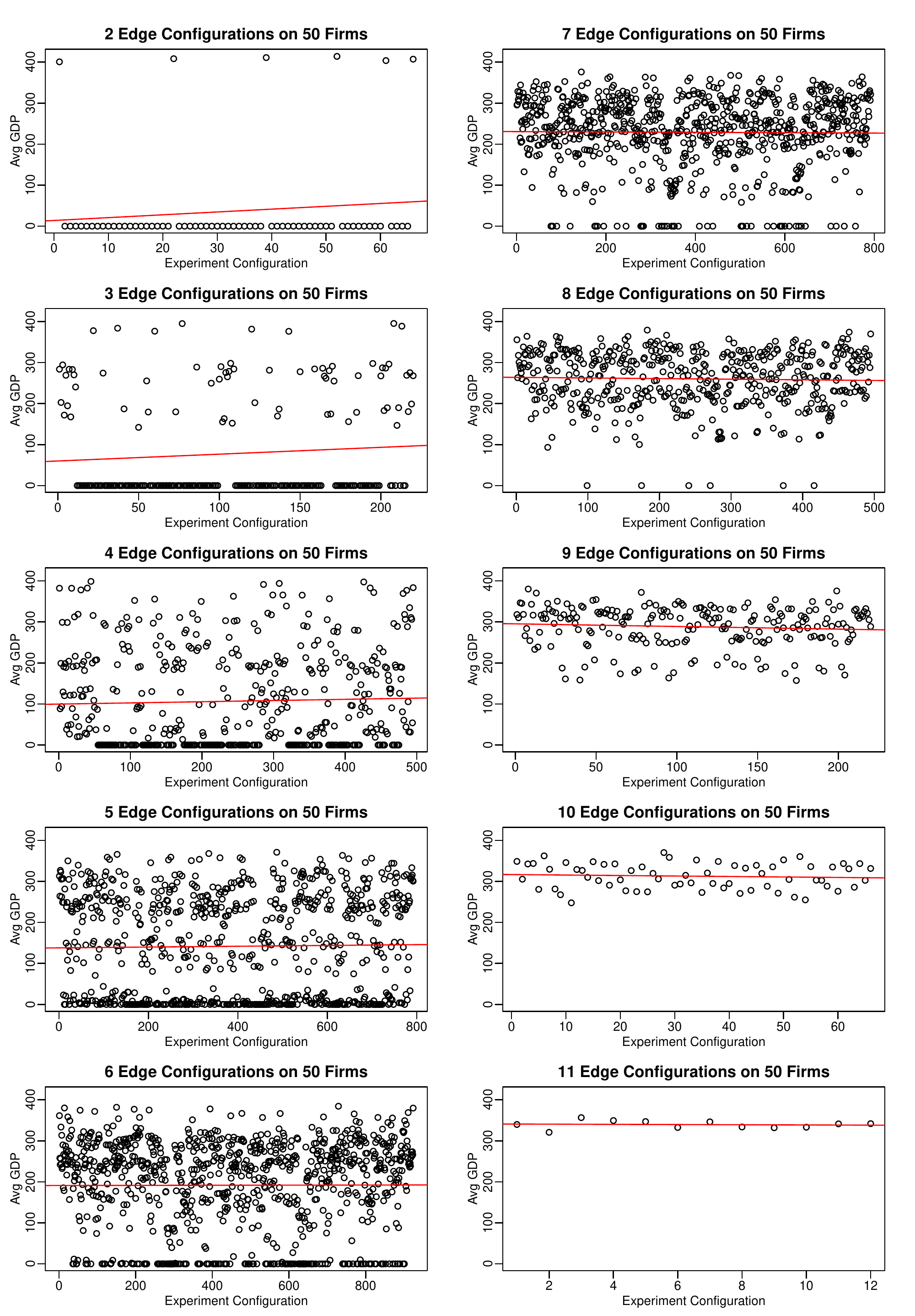}
\caption{Average GDP values for all edge configurations, by group, tested under a population of 50 firms. Each circle represents the average GDP over 20 replications of a specific experimental configuration. The line present in each plot is a regression line representing the average GDP value over all configurations in the specific edge group.}
\label{fig:raw_gdp}
\end{figure}

Figure \ref{fig:raw_gdp} shows an integrated picture of economic performance using the data from all 4095 subgraphs. When the number of edges is low, many configurations are unable to maintain economic performance due to a lack of cycle formation. This results in an average GDP of 0. However, as the edge count increases, particularly once it crosses from 6 edges to 7 edges, the number of graphs with cycles also increases and brings with it the ability for resources to be reused and recycled. This supports the idea that the reason for the increase in the average of average GDP is due to the ability of graphs with a higher density to form cycles, and may explain why the minimum GDP of the configuration sets jumps from 0 and begins to increase once a graph has more than 8 edges (figure \ref{fig:plots_b}).

\section{Discussion}

In this paper, we demonstrated that in a closed economic system with homogeneous fixed pricing and completely connected trade networks the density of the underlying resource transformation network has a significant effect on the economic performance of the associated economy. As the density of the resource transformation network increases so too does the average GDP of the economy (figure \ref{fig:plots_a}). However, an increase in the density of the transformation network also yields a slight decrease in the maximum potential GDP while at the same time produces a substantial increase in the minimum potential GPD (figure \ref{fig:plots_b}). Because the edges of the transformation network can be interpreted as technology, our results indicate as economies develop more technology, they can expect an increase in economic performance.

For economies with relatively small transformation networks, our findings suggest that it may be better to explicitly attempt to structure the network into its optimal configuration by restricting the ability to use specific technologies or create new edges through innovation and invention. However, such policy controls are not practical for large transformation networks because the number of subgraphs quickly becomes unmanageable and impractical to explore. In this case, when an economy is built on a large transformation network, it is better to foster innovation and technological adoption so as to maximize the expected economic performance through an increase in the average GDP, while at the same time increasing the minimum economic performance.

For transformation networks with a very large number of nodes, it may be physically or economically impossible to ever arrive at a state of complete connectivity. In this circumstance, our results suggest that as long as innovation can be maintained the economy should, on average, continue to improve its performance. This continued improvement in performance over time can be interpreted as economic growth.

\begin{figure}
\centering

\subfloat[]{\label{fig:plots_a}\includegraphics[width=58mm]{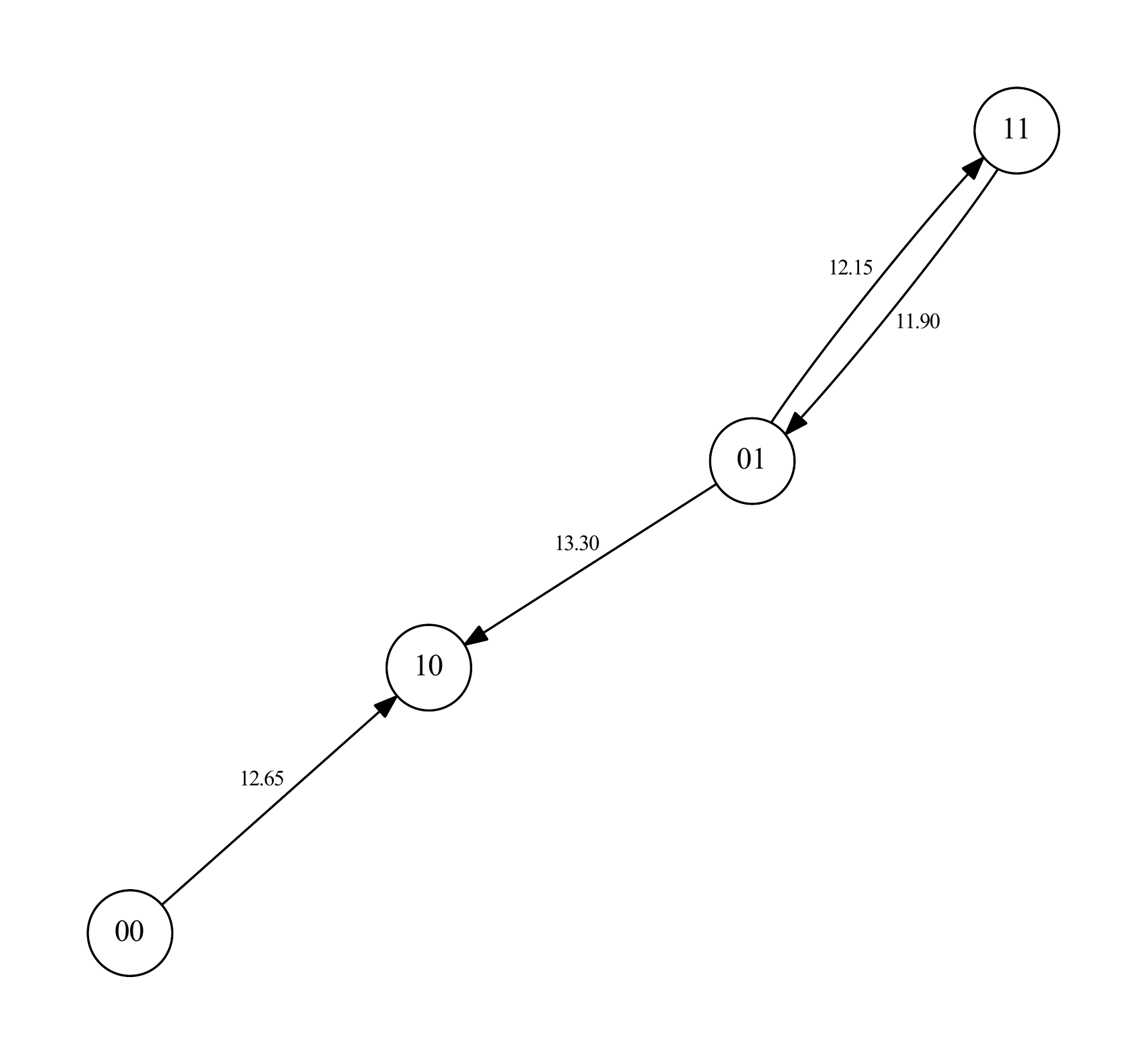}}
\hfill
\subfloat[]{\label{fig:plots_b}\includegraphics[width=58mm]{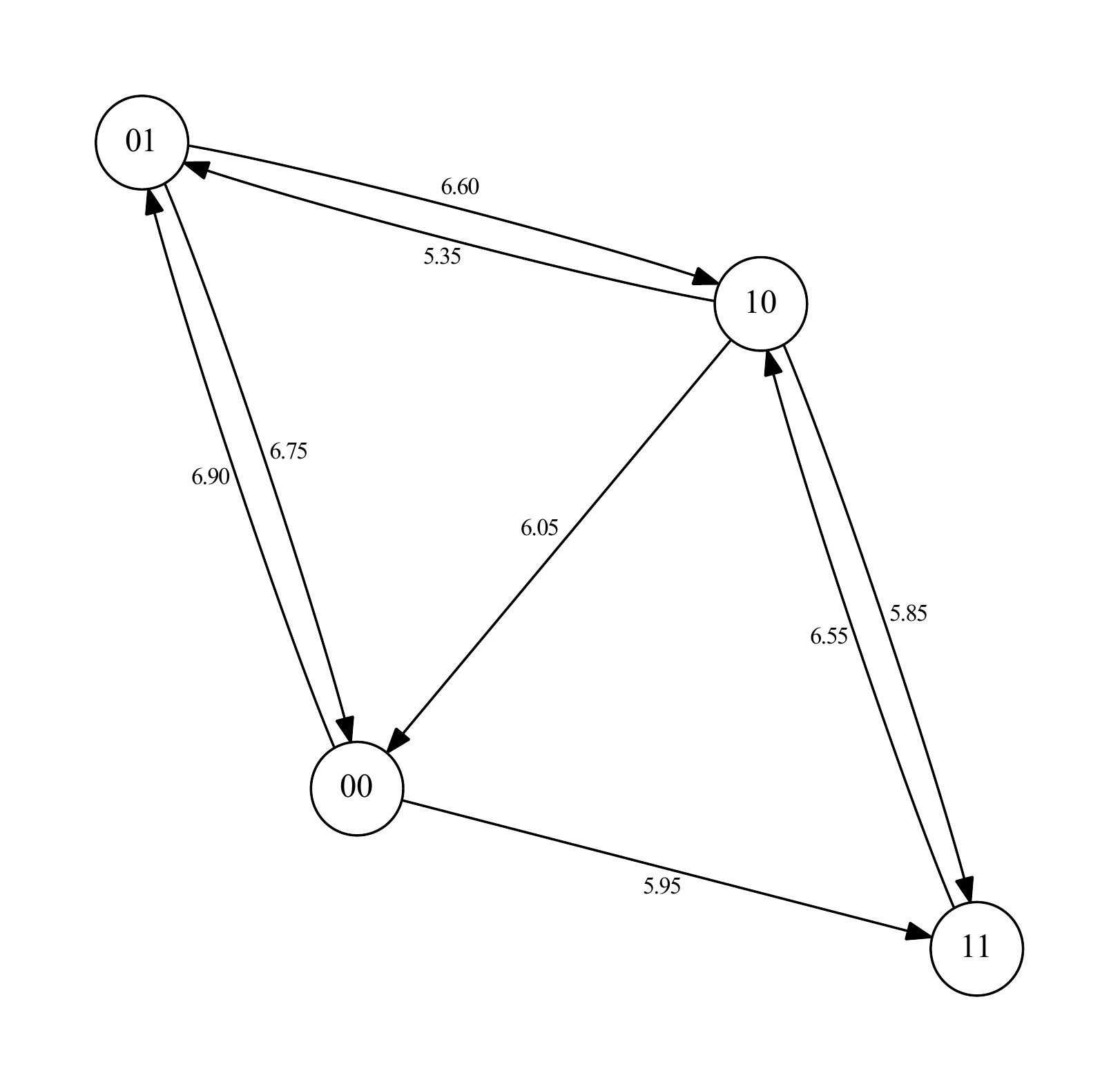}}

\caption{ (a) Graph configuration 651, composed of 4 nodes and 4 edges. This configuration yields an average GDP of 174.69 with 50 firms. (b) Graph configuration 3452, composed of 4 nodes and 8 edges. This configuration yields an average GDP of 355.89 with 50 firms. In both figures, the edge weights represent the average number of agents in the population with the associated transformation rule. For instance, in subfigure (a) 12.65 agents can transform resource 00 into resource 10. These weight values are determined by the distribution of transformation rules.}
\label{fig:graphs}
\end{figure}

However, it is not enough for an economy just to have resources, it must also have the capability and knowledge to use them. To illustrate this point, figure \ref{fig:graphs} presents two randomly chosen transformation networks from our data set. Both transformation networks use the same resources, but their edge configurations differ. This differences in edge configuration represents a difference in the available technology of the economy, and makes a significant difference in the realized performance. As a real world example, it is not enough to simply have a diamond mine in your country, you should also be able to turn the rough diamonds into jewelry. Likewise, not everybody should use the same technological processes. A healthy economy requires diversity.

\section{Future Work}

This paper has shown that looking at the transformation network of an economy may be a worthwhile way to gain insight into economic performance. In the future we plan to expand on this topic by adding incremental complexities to our model and comparing our results to real-world transformation networks where they can be identified and analyzed. 

Our next step is to examine the impact of transformation network density on GDP in larger, non-complete complex trade networks, such as scale-free, small world, and random topologies. We will then account for variable and heterogeneous decentralized pricing mechanisms. We hypothesize that the underlying trade network will have a larger impact on the economic performance of a system than the pricing, especially given the importance that cycles seem to play in a close system. 

Additionally, we need to investigate what happens if the system is not closed due to an external input of basic resources or an output of energy in the form of costs. Results from such investigations may have a significant impact on the understanding of cycles in a transformation network. We also need to look at economies with multiple transformation networks that may or may not be connected.

One of the more interesting results that we obtained also warrants further exploration. As shown figure \ref{fig:plots_b}, there appears to be a phase transition on the minimum GDP once the network obtains a sufficient density. This raises questions about the generality and calculation of this finding that should be explored in future work.

Finally, we plan to examine the properties of large transformation networks that evolve over time (dynamic networks instead of static networks). Because evolution can lead changes in the network structure by creating and destroy edges and altering path lengths, we expect that the more productive societies will have evolved a larger density in their underlying transformation networks.

\section{Conclusion}

Continued innovation and the amount of available technology are significant factors in economic growth. 

With the world in economic crisis, new perspectives on economics are needed. Transformation networks provide one new way to look at economies by examining how the system's underlying economic resources are transformed from one form into another. This is a general approach that can be used on any economic sector or within any industry, and more generally to any system in which one thing is changed into another by some entity. 

We have shown that these networks are not just an abstract idea. Their structure can have real impacts on economic performance. In particular, an increase in the density of a transformation network appears to cause an increase in both the average GDP of the associated economy and an increase in the minimum GDP. This increase in density can be associated with innovation in the economy. As firms discover new ways to use the resources they have available, and invent new ways to turn one resource into another, new edges are added to the underlying network.

Our results suggest that in economies that are small and simple, the transformation networks can be fully explored and optimal network configurations identified. However, when economies become even moderately large in the number of resources they rely upon, such explorations are impractical and innovation must be relied upon to increase growth.

This paper presents the first of many steps towards understanding and leveraging transformation networks; but even within the relatively simple approach that we have taken it has been shown that these networks can provide key insights with regard to explanations of economic performance and potential areas of innovation.

\bibliographystyle{spmpsci}
\bibliography{references}

\end{document}